# Metal-less Optical Left Handed Material by Low-dimensional Quantum Structure Anisotropy

Pavel Ginzburg[1] and Meir Orenstein[1]

*[1] Department of Electrical Engineering, Technion, Haifa 32000, Israel*

**Left-handed materials (LHM) are artificial composite materials having the unique property of exhibiting negative index of refraction [1]. These negative-refraction materials refract light in a way, which is contrary to the normal "right handed" rules of electromagnetism. The hope is that the peculiar properties will lead to superior lenses [2], and may provide a possibility to observe negative analogies of other prominent optical phenomena, such as the Doppler shift [1] and Cerenkov radiation [1]. Most of nowadays proposed and demonstrated LHM metamaterials designed for the visible spectrum are based on metal inclusions which cause an extremely high energy loss which limits of their applicability [3, 4, 5, 6, and 7]. Here we propose a novel concept of metamaterial assembly where metal inclusions are replaced by semiconductor based low-dimensional quantum structures, exhibiting significantly lower losses and even may be inverted to exhibit gain by carriers' injection. Another theoretical realization of gaseous non-metallic LHM [8] employs a quantum interference technique (EIT) [9] and suffers from the high density of gas atoms and the ultranarrow frequency range of the LHM properties. We demonstrate, how to overcome these disadvantages, exhibiting relatively wide band LHM, combined with ease of preparation – namely growing low-dimensional semiconductor quantum wells (QW) [10] and quantum dots (QD) [11]. One of the most important advantages of the proposed scheme is the ability to invert the material losses to gain by using active, pumped quantum structure – a virtue absent in metals.**

We start from a configuration recently proposed in [4] and further developed in [12], where the negative µ is replaced by an exceptionally highly anisotropic waveguide structure. The transverse magnetic (TM) optical modes, propagating within the waveguide with positive-negative anisotropic core, will be effectively left-handed. We demonstrate two possible structures consisting of QWs or coupled QDs to achieve the required anisotropy ($\varepsilon_\perp < 0$, $\varepsilon_\parallel > 0$), and thus to implement a LHM. The waveguide core will contain the layered QWs (Fig.1 (a)) or vertically grown aligned QDs stack (Fig.1 (b)). The anisotropy in our scheme originates from the difference between the inplane and vertical (growth direction) susceptibility of the quantum structure for the intersubband transitions, namely dipolar selection rules for QWs and polarization dependent tunnelling efficiency for coupled QDs. Although the pronounced effects are in the vicinity of transition resonances, much lower loss relatively to metal based metamaterials is expected. Additional merit of the quantum structure is a tuning capability (e.g. from LHM to RHM), by applying voltage on the structure, and, probably, the most outstanding feature – gain by carrier injection is possible for semiconductor LHM realizations using the quantum cascade amplifier configurations [13, 14].

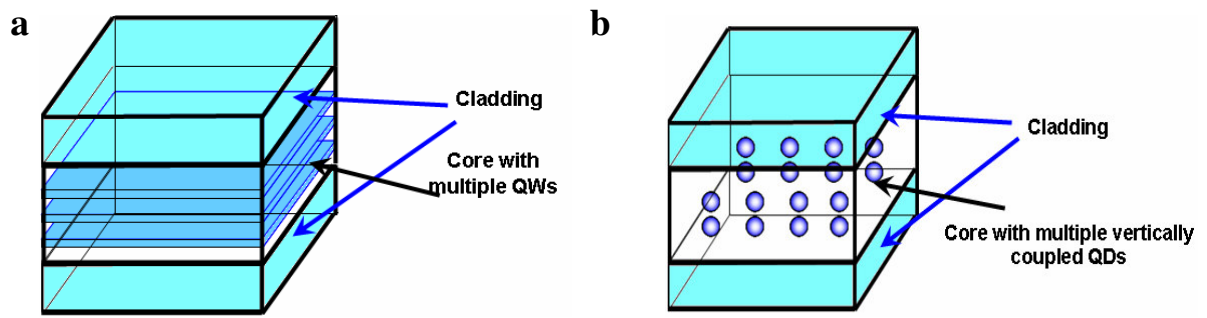

Fig.1 – Slab waveguides with anisotropic core composed from: (a) quantum wells (b) coupled quantum dots.

For the generation of the particular anisotropy, later to be translated to negative refractive index, we used GaAs/AlGaAs heterostructure. Actually, the GaAs/AlGaAs and InGaAs/AlInAs composites typically used for infrared detectors and quantum cascade lasers are well established both for their material properties and technological processing. The only disappointing feature for our purpose disadvantage is that these material groups have relatively small conduction band (CB) offset which locate the intersubband optical transitions within the infrared region. However, novel nitride alloys [15] - actually InN/GaN/AlN QWs/QDs [16, 11] are exhibiting a large CB offsets and wide energy gap, enabling near IR and visible range interactions via intersubband transitions. In addition, the Fermi level in such nitride composite is located within the CB which allows the passive employment of these structures (with no electrical pump or doping). However – the material properties and spectroscopic data of this material family is still incomplete

The origin of optical permeability tensor anisotropy (Eq.1) in QWs is the polarization dependent dipole matrix element for intersubband transitions. The light beam being polarized in the plane of the QWs growth will strongly interact with the resonant CB electron, while the polarization perpendicular to the growth direction will not be affected from the QW.

$$\varepsilon = \begin{pmatrix} \varepsilon_\perp & 0 & 0 \\ 0 & \varepsilon_\parallel & 0 \\ 0 & 0 & \varepsilon_\parallel \end{pmatrix}, \varepsilon_\parallel = \varepsilon_{back} > 0, \varepsilon_\perp < 0 \quad (1)$$

The intersubband induced permeability can be represented by Lorenzian model [16]:

$$\varepsilon_\perp = \varepsilon_{bac} - iNq \frac{|\mu_{12}|^2 (\rho_{11} - \rho_{22})}{\varepsilon_0 \hbar} \frac{1}{i(\omega - \omega_{21}) - \gamma_{21}} \Gamma \quad (2)$$

where q is the electron charge, $\varepsilon_0$ is the vacuum electric permeability, $\omega_{12}$ is the resonant intersubband transition frequency, $\omega$ is the central frequency of the input light, $\rho_{ii}$ is the electron occupation density of the i-th level, (reduced to the ground state for passive device and to the exited for active), N is the density of the ground state electrons, $\varepsilon_{bac}$ is the averaged permeability of the background material, γ is phenomenology introduced dephasing rate and Γ is the optical field confinement factor within the well layer. Theoretically, by choosing the appropriate parameters, it is possible to reach the negative real part of ε in the vicinity of transitional resonance, but practically there are number of bottlenecks to overcome. The first is to achieve a sufficiently high ground state electron density by strong electrical pump or by doping of QW layer; nevertheless the electron population still will be limited by $\sim 10^{19} cm^{-3}$ due to the recombination processes. The outstanding advantage of nitrides is the location of the room temperature Fermi level, lying within the CB which is contributing to very high electron population (up to the Avogadro number!).

The other crucial parameter is the dephasing. The dephasing rate is in general a consequence of several physical processes. The most significant contributors are interface roughness scattering, phonon scattering, impurity scattering, alloys disorders, many body effects (electron-electron, electron-hole scattering) [17]) and subband dispersion [18]. While the phonon scattering may be significantly reduced by temperature manipulation, the further reduction of the dephasing rate is limited by technology issues [17] as well as the intrinsic material structure [18]. The nowadays dephasing values are about few meV for cryogenic QWs and tens of meV for room temperature devices. For nitride heterostrucures this numbers are still larger [19] due to the immaturity of their technology, however the theoretical values are much more promising and are better by an order of magnitude compared to other existing materials [20].

The last factor corresponding to optical permeability is the field confinement $\Gamma$ (Eq.1), which is relatively small in conventional structures of QW superlattices and lasers, used in the spectroscopy experiments. This is the reason, why the very deep resonances were measured but yet without demonstrating the negative dielectric response [21]. The very recent work of the C. C. Phillips group [22] exhibiting negative resonances in a semiconductor inversionless laser removes the doubts that negative $\varepsilon$ can be achieved in the vicinity of QW intersubband resonance.

We explored the well established heterostructure $GaAs/Al_{0.3}Ga_{0.7}As$ as a simple example to demonstrate a medium exhibiting the negative $\varepsilon$. Many other variants of material families will exhibit the same phenomena and can be optimally engineered for applicable devices. The parameters for $GaAs/Al_{0.3}Ga_{0.7}As$ QW were taken from experimental data [16]: well depth - $\Delta E_c$ = 230meV, well width L = 10nm, the volume density of charge $1 \cdot 10^{18} cm^{-3}$. Dephasing rates of the dipolar transitions were chosen to be $\gamma_{12}$ =3meV [17] $\varepsilon_{back}$ is the average between the constituents and its value is ~13. The waveguide core is designed to contain about 30% QW layers. The simulation results depicted in Fig.2 (a) exhibit evident negative resonance in the far infrared range. The simulation neglects the free carrier absorption losses that are negligible compared to the resonant transition loss. Choosing the operation wavelength to be 13μm we obtain the $\varepsilon_\perp$ = -1.56 + 5.22j, while $\varepsilon_\parallel$= 13. For a 13μm slab embedded in perfectly conducting layers ([4] configuration) we estimate the effective refractive index for the 4[th] TM mode to be -0.9+0.6j and for realistic situation slab in the air [12] the 5[th] TM mode gives -2.1+1.6j refractive index. In spite of the high loss this result is much better than the recently reported results based on metallic inclusions [7], and may be further improved at cryogenic temperature or using the other material alloys.

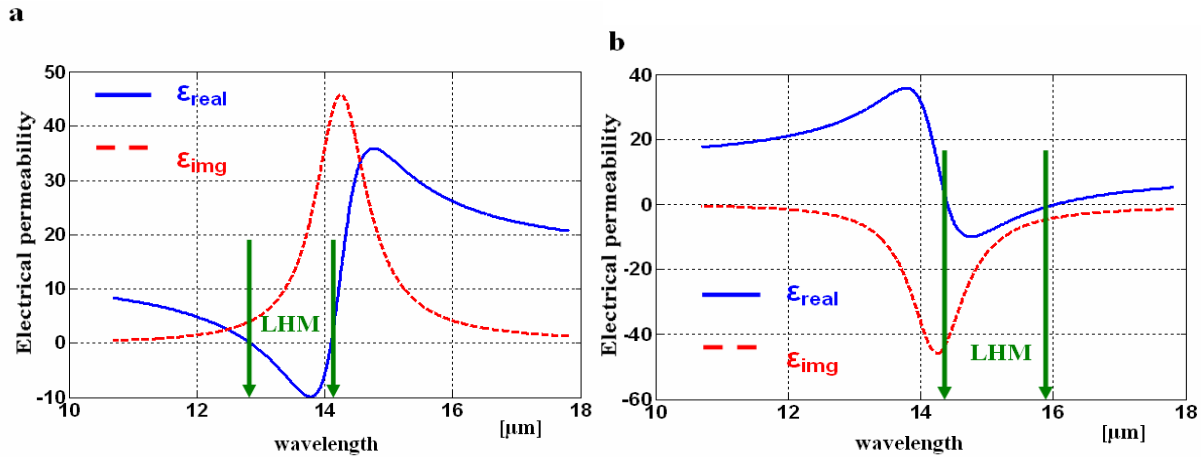

Fig.2 – Perpendicular term of electrical permeability tensor of the waveguide core. Green arrows highlight the LHM region: (a) passive device (b) active device.

Additional significant feature of the proposed configuration is its controllability. The control may be achieved, by coupling QWs and applying the inverse bias voltage on them and switching the material regime from LHM to RHM. The detailed analysis for this configuration exists in our resent work [23]. Exploiting the configuration of [23] we depict in Fig. 3 the voltage switching between LHM and RHM regimes. This is performed by voltage control of the $\varepsilon\bot$ sign. For zero bias, the e value is negative at the prescribed wavelength, while, for a large bias the coupled QW move out resonance and the permeability move to the positive – RHM regime. Furthermore, introducing a structure of multiple coupled wells, namely a QWs superlattice, not only enhances the confinement factor of the optical mode, but significantly increases the operation bandwidth of the device [29].

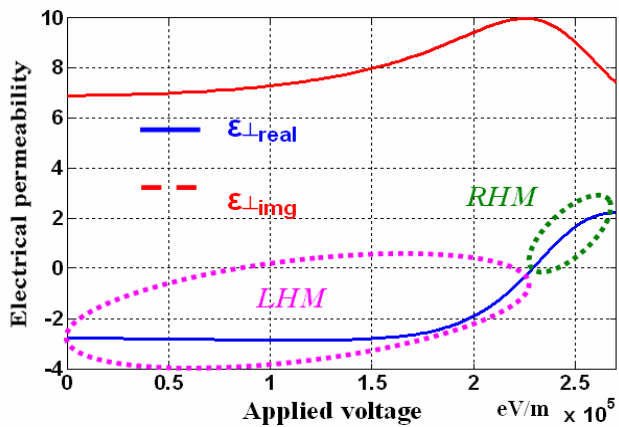

Fig.3 – Voltage controllability of material regime. Magenta ellipse highlighted the LHM region; green ellipse highlighted the RHM region.

The ultimate advantage of semiconductors over metals is a possibility to convert the loss to gain by electrical current pump, inverting the population, while preserving the previously discussed positive negative anisotropy. For the intersubband transitions the configuration of an active region of a quantum cascade laser will serve as the gain LHM slab. Actually the expression for $\varepsilon$ (Eq. 1) is still valid when replacing the expression $\rho_{11} - \rho_{22} \approx 1$ to be $\rho_{11} - \rho_{22} \approx -1$ (complete population inversion). This can be achieved in a conventional quantum cascade lasers by introducing additional (third) energy level with fast nonradiative transition rate and further tunnelling to the next cascade level. Under the same discussed constrains and injecting electron current, an active LHM may be achieved (Fig.2 (b)). The resulting parameters are $\varepsilon_\perp = -9.9 - 24.5j$, $\varepsilon_\parallel = 13$ leading to refractive index of $-3.4 - 2.3j$ for 13μm slab embedded in perfectly conducting layers. The 4$^{th}$ TM mode of the slab in the air [12] leads to $-2 - 1.47j$.

The realization possibility of the proposed QWs based on anisotropic slab waveguide is quite evident and its demonstration is feasible by nowadays technology. However, significant losses still exist for the passive configuration. Here we propose theoretically a passive device based on QDs with significantly reduced losses, seemingly being the limit for the proposed technique of quantum structures embedded metamaterials. The usage of QDs is more promising because the three-dimensional carrier confinement significantly reduces the dephasing rates [24]. Self-assembled QDs [11] has a prominent size distribution resulting in an inhomogeneous line broadening [25], which may currently prevent the implementation of the LHM. However controlled growth of ordered QDs is now under intense research effort [26] and, hopefully will serve the solution for QDS based LHM.

For the demonstration of QDs based LHM, the strongly anisotropic structure with negative resonances will be constructed from pairs of vertically coupled InN/AlN QDs [27]. For simplicity of the numerical calculation we assume the QDs are spherical in shape, but the existing hexagonal- pyramidal InN/AlN QDs [11] will lead the similar results. It should be noted that realistic singular QDs are anisotropic structures [30] but we decided to exploit the coupled QDs configuration to achieve the controllability as discussed in the previous section. Coupled pairs of 3nm diameter QDs can be prepared by growing two layers of aligned QDs, separated by 5nm. We select the spacing between the dots to provide a finite tunnelling [27] for the excited states (p-states) while negligible for the ground state (s-state) (explicit calculations give a ~53 coupling ratio). The Fermi level crossing the CB will promise the electron occupation of the s-shaped ground state. The "bonding" due to the exited states coupling depends on the angular functions due to conservation of the angular momentum in a resonant tunnelling process [28]. In Fig. 3 the coupling of the wavefunctions with angular momentum along (Fig. 4(a)) or transverse (Fig. 4(b)) to the molecular axis (growth direction) is shown schematically. It provides an intuitive explanation for the origin of anisotropy.

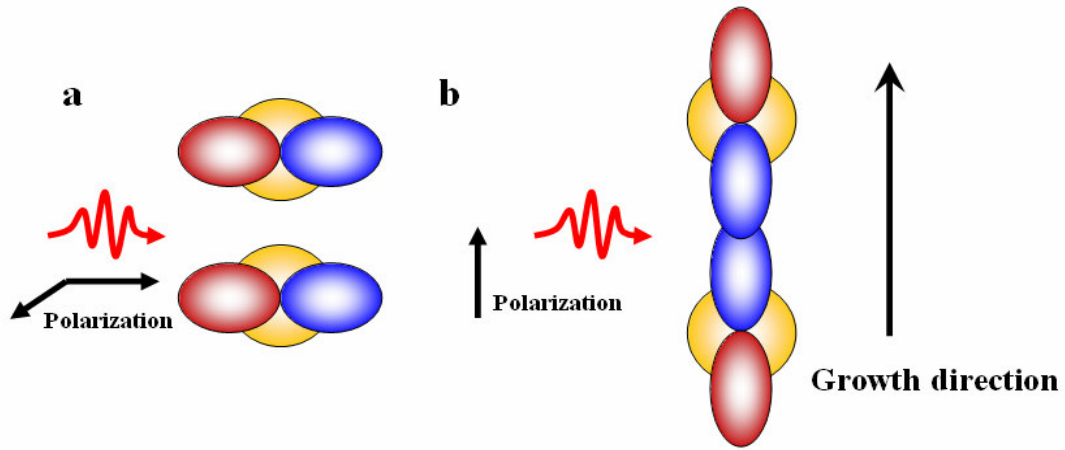

Fig.4 – Schematics of metamaterial comprised of coupled QDs. Dipole selection rules define the orientation of the p-orbital of spherical QD. Vertically polarized incident light interacts with highly coupled QDs, while horizontal polarization encounters negligibly coupled structure.

The explicit calculations give a ratio of ~15.4 between the respective coupling constants. The detailed treatment of such four-level system may be found in [23]. Strong anisotropy is not enough for realizing LHM. The medium should exhibit also negative horizontal and positive vertical ε values. In order to address this point we performed simulations of a layered structure comprised of coupled QDs, using the following material parameters: $\Delta E_c$=3.65eV, $m_{InN}$=0.07$m_0$, $m_{AlN}$=0.32$m_0$, $\lambda_{central}$= 965.5 nm, $\gamma$ =7 µeV, dots density N=4·$10^{23}$ m$^{-3}$. In the region indicated by an arrow in Fig.5, the Podolskiy-Narimanov criterion [4] $\varepsilon_\perp$<0 $\varepsilon_\parallel$>0 is fulfilled. The resulting values for electrical permeability at the wavelength of 956.3 nm, exhibiting the minimal material loss, are: $\varepsilon_\perp$ = -3.5+0.2j $\varepsilon_\parallel$ = 7.4+0.2j. The resulting refractive index is -5-0.01j for the 6$^{th}$ mode of 1µm slab imbedded in perfect conductors.

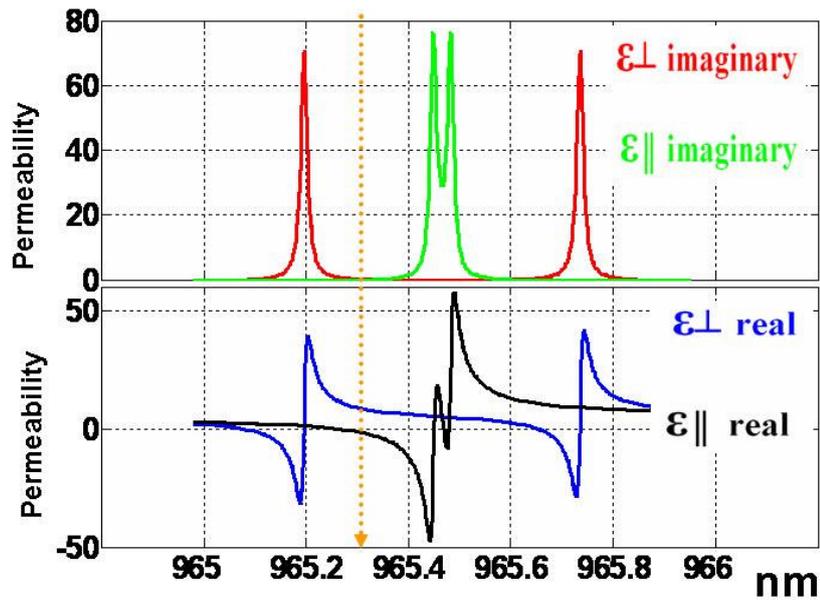

Fig.5 -Simulation of real and imaginary parts of electric permeability of the system of coupled QDs. The electric permeability curves represent the anisotropy of the ε tensor. The brown arrow on the graph points to the region minimal losses where Podolskiy-Narimanov criterion ($\varepsilon_\perp<0$ $\varepsilon_\parallel>0$) fulfilled – hence, this is the region of LHM.

In conclusion we proposed for the fist time the usage of low-dimensional quantum system for LHM implementation. Two different schemes: QWs and coupled QDs based anisotropic waveguides have been proposed. A significant loss reduction compared to the currently discussed LHM is estimated. Two significant features of gain and tunability, missing in existing metal based implementation are possible in our configuration. In addition, a fast development of nitride group materials shows a remarkable promise for realization of our proposition for visible light!

# References:

1. Veselago, V. G., "The electrodynamics of substances with simultaneously negative values of permittivity and permeability", Sov. Phys. Usp. **10**, 509–514 (1968)

2. Pendry, J. B. "Negative refraction makes a perfect lens", Phys. Rev. Lett. **85**, 3966– 3969 (2000)

3. V.A. Podolskiy, A.K. Sarychev and V.M. Shalaev "Plasmon modes and negative refraction in metal nanowire composites" – Optics Express **11,** 735 (2003)

4. V.A. Podolskiy and E.E. Narimanov "Strongly anisotropic waveguide as a nonmagnetic left-handed system" - Phys. Rev. B, **71** 201101(R) (2005)

5. Panina, L. V. , Grigorenko, A. N. & Makhnovskiy, D. P.," Metal-dielectric medium with conducting nanoelements",  Phys. Rev. B **66**, 155411 (2002)

6. Yen, T. J. W. J. Padilla, N. Fang, D. C. Vier, D. R. Smith, J. B. Pendry, D. N. Basov, X. Zhang, " Terahertz magnetic response from artificial materials", Science **303**, 1494–1496 (2004)

7. A. N. Grigorenko, A. K. Geim, H. F. Gleeson, Y. Zhang, A. A. Firsov, I. Y. Khrushchev, J. Petrovic, " Nanofabricated media with negative permeability at visible frequencies", Nature **438**, 335 - 338 (17 Nov 2005)

8. S. Jian-qi, R. Zhi-chao, H. Sai-ling, "How to realize a negative refractive index material at the atomic level in an optical frequency range?", Journal of Zhejiang University SCIENCE , Vol. **5**   No. 11   p.1322-1326 (2004).

9. S.E. Harris, "Electromagnetically Induced Transparency", Physics Today, **50** (7), pp. 36-42 (1997).

10. S. Ochi, N. Hayafuji, Y. Kajikawa, K. Mizuguchi and T. Murotani, " Growth of GaAs/AlGaAs quantum well structures using a large-scale MOCVD reactor",  Journal of Crystal Growth Volume **77**, Issues 1-3 , pp 553-557, (1986).

11. S. Ruffenach, B. Maleyre, O. Briot, and B. Gil, "Growth of InN quantum dots by  MOVPE", phys. stat. sol. (c) 2, No. **2**, 826–832 (2005).

12. Yinon Satuby, Noam Kaminsky, Meir Orenstein, "Nano Optical Modes in Gaps within Left-Hand-Metamaterial Waveguide," submitted to JOSA A Feature issue on "Photonic Metamaterials: from Random to Periodic" to be published (2007).
    Noam Kaminsky, Yinon Satuby, and Meir Orenstein, "Nano Optical Modes of Gap Structure in a Left-Hand-Metamaterial Waveguide", in Photonic Metamaterials: From Random to Periodic on CD-ROM (The Optical Society of America, Washington, DC, 2006), presentation number: WD11

13. Jerome Faist, Federico Capasso, Deborah L. Sivco, Carlo Sirtori, Albert L. Hutchinson, and Alfred Y. Cho, "Quantum Cascade Laser", Science **264**, pp. 553 – 556 (1994).

14. Carlo Sirtori, Peter Kruck, Stefano Barbieri, Philippe Collot, Julien Nagle, Mattias Beck, Je´roˆme Faist, Ursula Oesterle, "GaAs/AlxGa1-xAs quantum cascade lasers", Appl. Phys. Lett. **73** 3486-3488 (1998).

15. B. Monemar and G. Pozina, "Group III-nitride based hetero and quantum structures", Progress in Quantum Electronics **24**, 239-290 (2000).

16. Myung Goo Cheong, E-K Suh and H J Lee, "High-quality $In_{0.3}Ga_{0.7}N/GaN$ quantum well growth and their optical and structural properties", Semicond. Sci. Technol. **16** (2001) 783–788

16. P. Basu, Theory of optical processes in semiconductors: bulk and microstructures, (Oxford: Clarendon Press, c1997).

17. K. L. Campman, H. Schmidt, A. Imamoglu, and A. C. Gossard, " Interface roughness and alloy-disorder scattering contributions to intersubband transition linewidths", Applied Physics Letters, **69**, pp. 2554-2556, (1996).

18. Inès Waldmüller, Michael Woerner, Jens Förstner, Andreas Knorr, "Theory of the lineshape of quantum well intersubband transitions: optical dephasing and light propagation effects", physica status solidi (b), **238,** pp. 474 – 477, (2003).

19. J. Hamazaki, H. Kunugita, and K. Ema, A. Kikuchi and K. Kishino, "Intersubband relaxation dynamics in GaN/AlN multiple quantum wells studied by two-color pump-probe experiments", Phys. Rev. B **71**, 165334 (2005).

20. W. Chow, M. Kira, S. W. Koch, "Microscopic theory of optical nonlinearities and spontaneous emission lifetime in group-III nitride quantum wells", Phys. Rev. B **60**, 1947 - 1952 (1999).

21. K. Unterrainera, R. Kerstinga, R. Bratschitscha, G. Strassera, J. N. Heymanb, K. D. Maranowskic and A. C. Gossardc, "Few-cycle THz spectroscopy of nanostructures", Physica E: Low-dimensional Systems and Nanostructures, **7**, pp. 693-697(2000).

22. M. D. Frogley, J. F. Dynes, M. Beck, J. Faist and C. C. Phillips, "Gain without inversion in semiconductor nanostructures", Nature Materials **5**, 175–178 (2006).

23. P. Ginzburg and M. Orenstein, "Slow light and voltage control of group velocity in resonantly coupled quantum wells," Opt. Express **14**, 12467-12472 (2006).

24. N. H. Bonadeo, J. Erland, D. Gammon, D. Park, D. S. Katzer and D. G. Steel, " Coherent Optical Control of the Quantum State of a Single Quantum Dot", Science. **282.** pp. 1473 – 1476 (1998).

25. J. Misiewicz, G. SImagek and K. Ryczko, "Photoreflectance spectroscopy of quantum dots", Current Applied Physics **3**, pp. 417-420 (2003).
26. J. H. Lee, Zh. M. Wang, B. L. Liang, K A Sablon, N. W. Strom and G. J. Salamo, "Size and density control of InAs quantum dot ensembles on self-assembled nanostructured templates", Semicond. Sci. Technol. **21**, 1547–1551 (2006)
27. E. A. Stinaff, M. Scheibner, A. S. Bracker, I. V. Ponomarev, V. L. Korenev, M. E. Ware, M. F. Doty, T. L. Reinecke and D. Gammon, "Optical Signatures of Coupled Quantum Dots", Science **311**, 636 (2006).
28. U. Gennser, M. Scheinert, L. Diehl, S. Tsujino, A. Borak, C. V. Falub, D. Grützmacher, A. Weber, D. K. Maude, G. Scalari, Y. Campidelli, O. Kermarrec and D. Bensahel, "Total angular momentum conservation during tunnelling through semiconductor barriers", Europhys. Lett., **74** (5), pp. 882-888 (2006)
29. Pavel Ginzburg and Meir Orenstein, "A periodic structure of coupled double quantum wells for significant light slowing", in Photonic Metamaterials: From Random to Periodic on CD-ROM (The Optical Society of America, Washington, DC, 2006), presentation number: ThD19
30. Mitsuru Sugisaki, Hong-Wen Ren, Selvakumar V. Nair, Kenichi Nishi, Shigeo Sugou, Tsuyoshi Okuno, and Yasuaki, " Optical anisotropy in self-assembled InP quantum dots", Phys. Rev. B **59**, R5300 (1999)